\documentstyle[12pt,texdraw]{article}

\begin{document}

\title{\bf On the large order behaviour of the Potts model}
\author{Domenico Maria Carlucci \\
        Scuola Normale Superiore \\ 
         Piazza dei Cavalieri 7 \\  
         56100 Pisa, Italy \\ 
        {\small Internet: {\tt CARLUCCI@UX2SNS.SNS.IT}} \\
	{\small Fax number: +39 50 563513}  
        }

\date{\today}

\maketitle

\thispagestyle{empty}   

\vspace{3cm}

\centerline{
{\bf Pacs numbers:} {05.20.-y, 11.10.-z, 11.10.Jj}
}

\centerline{
{\bf Section proposed:} General Physics
}

\clearpage

\thispagestyle{empty}   

\begin{abstract}
Following the work by Houghton, Reeve and Wallace about an 
alternative formulation of the $n \to 0 $ limit of the 
$(n+1)$ state Potts model in field theory for the large order behaviour of the 
perturbative expansion, 
we generalise their technique to all $n$ by establishing an 
equivalence in perturbation theory order by order with another 
bosonic field theory. Restricting ourselves to a cubic interaction, 
we obtain an explicit expression (in terms of~$n$) 
for the large order behaviour of the partition function.
\end{abstract}

\vspace{3cm}

\begin{center} 
    {\bf Resum\'e} 
\end{center}
Suivant le travail de Houghton, Reeve et Wallace concernant 
une formulation alternative de la limite $n\to 0$ du 
mod\`ele de Potts en th\'eorie  des champs pour le comportement 
aux grands ordres, nous g\'en\'eralisons leur 
technique \`a $n$ quelconque en \'etablissant une 
\'equivalence ordre par ordre en th\'eorie de 
perturbation avec une autre th\'orie de champs de bosons. 
Nous restregnant \`a une interaction cubique, nous obtenons une 
expression explicite (en fonction de $n$) pour le comportement 
aux grands ordres de la fonction de partition.

\clearpage

\section*{Introduction}

The Potts model\cite{Potts} is one of the most fascinating 
topics in statistical mechanics. It can be  regarded as the 
generalisation of the Ising model\cite{Ising} to a generic 
number $q$ of components. Originally this problem was proposed 
to Potts by his PhD Professor Domb\cite{Domb}, who suggested him 
to investigate the critical properties of a system of interacting 
spins confined  in a plane, pointing in $q$ equally spaced 
directions 

	\begin{equation}
           \Theta_i\,=\,\frac{2\pi i}{q}
	   \hspace{1.5truecm}
	   i=0,1,...,q-1  
	\end{equation}
with an hamiltonian depending only on the relative orientation of two 
neighbouring spins of the form 

	\begin{eqnarray}
	   {\cal H} =
	   - \sum_{\{\rm spins\}}
	   J(\Theta_i,\Theta_j) 
           & \,\,\, &
	   J(\Theta_i,\Theta_j) \propto \cos(\Theta_i - \Theta_j)
	\end{eqnarray}

Actually, the previous model bears the name of {\it planar Potts model}.
Nevertheless, the most famous model is the one where two spins 
interact only when they are present in the same state, that is 

	\begin{equation}
           J(\Theta_i,\Theta_j)\,\propto\,
	   {\delta}_{\Theta_i,\Theta_j}^{\mbox{\tiny (Kr)}}.
	   \label{real_Potts_model}
	\end{equation}
The interaction (\ref{real_Potts_model}) can be re-written 
in  such a way as to reflect a full-symmetry in a $(q-1)$-dimensional 
space by making use of $q$ vectors $(q-1)$-dimensional 
$\mbox{\boldmath $\epsilon$}^{\alpha}$ pointing in the $q$ symmetric
 directions of a hypertetraedron in $(q-1)$ dimensions.
Indeed, the interaction (\ref{real_Potts_model}) can be re-written as  

	\begin{equation}
	  \delta(\Theta_i,\Theta_j)\,=\,
	  \frac{1}{q}\,\left[
                             \,1\,+\,(q-1)\,
	                     \mbox{\boldmath $\epsilon$}^{\Theta_i}
			     \cdot
			    \mbox{\boldmath $\epsilon$}^{\Theta_j}
		       \right].
	\end{equation}
The increasing attention to the Potts model was due to the fact that it has 
proven to be related to a large class of outstanding problems, especially 
when particular values of $q$, even non integer, are taken. 
For example, the percolation problem can be formulated in terms of $q=1$ Potts 
model\cite{Kasteleyn,Giri,Kunz}.  
Also Fortuin and Kasteleyn\cite{Fortuin} showed that the Kirchhoff's 
solution\cite{Kirchhoff} for an ensemble of resistors 
is strictly related to $q=0$ limit of the 
Potts model.  Moreover, the critical behaviour of the dilute 
spin glasses is well understood in terms of the $q=\frac{1}{2}$ 
limit\cite{Aharony1,Aharony2}. Finally, Lubensky and Isaacon 
\cite{Lubensky} showed that interesting processes in gelation 
and vulcanization fall in the same class of universality of 
$0\leq q \leq 1$ Potts model (For a complete review of the Potts model 
see F.Y.Wu\cite{Wu}).

A convenient formulation of Potts model as a field theory was first 
performed by Zia and Wallace\cite{Zia}, who investigated 
the critical properties in the framework of the renormalization 
group. Actually in the field theory literature 
it is customary to refer to $(n+1)$ Potts 
model ($n+1=q$ in the previous notation).  
The symmetry of the model allows a trilinear interaction which is the 
leading term in the renormalization group framework.  
Because of a pure cubic interaction, the high order behaviour  of the 
perturbative expansion is clearly non-oscillating and therefore non 
Borel-sommable. Nevertheless, 
when particular limits are taken ({\it e.g.} $n\to 0$ percolation 
problem) the large order behaviour is found to be oscillating and 
Borel-sommable\cite{HRW}) as obtained by Houghton, Reeve and 
Wallace (from now on referred as HRW). 
In  Sec.I, we recall the hamiltonian of the $(n+1)$ Potts model 
and the  
alternative approach to $n=0$ limit used by Houghton, Reeve and Wallace.
In Sec. II its generalisation to each value of $n$ is showed. 
Sec.III will be devoted to the evaluation  of the 
large order behaviour for such an alternative field theory and the results 
will be shown to agree with the previous ones.

\section{The $(\lowercase{n}+1)-$state Potts model} \label{Potts_model} 

We consider the $(n+1)$ state Potts model in 
$d$ dimension, described by an \\ Hamiltonian\cite{Zia} of the form

	\begin{equation}
	  {\cal H}[\mbox{\boldmath $\phi$}]=
	   \int
	   d^d x
	         \left[
		        \frac{1}{2}
			(\nabla \phi_i)(\nabla \phi_i)
			\,+\,
			\frac{1}{2}\phi_i\phi_i
			\,+\,
			\frac{g}{3!}
			\rho_{ijk}\phi_i\phi_j\phi_k
		\right]
		\label{H_Potts}
	\end{equation} 
where sum over repeated indices is implied and  
runs from $1$ to $n$. The system possesses a hypertetrahedric 
symmetry, provided that $\rho_{ijk}$ is of the form

	\begin{equation}
	  \rho_{ijk}=\sum_{\alpha=1}^{n+1} 
	             \epsilon_i^{\alpha}
	             \epsilon_j^{\beta}
		     \epsilon_k^{\gamma}
	\end{equation} 
where the $\epsilon$'s satisfy

	\begin{equation}
		\sum_{\alpha=1}^{n+1}\epsilon_i^{\alpha}= 0   
                \,\,\,\,\,
	        \sum_{\alpha=1}^{n+1} 
                                      \epsilon_i^{\alpha}
				      \epsilon_j^{\alpha}
                                      = (n+1)\delta_{ij} 
		\,\,\,\,\,
                \sum_{i=1}^{n+1}\epsilon_i^{\alpha}
					  \epsilon_i^{\beta}= (n+1)
					  \delta^{\alpha\beta}-1
		\label{epsilon_formulas}    
         \end{equation}
Physical quantities can be extracted from the partition 
function (the discussion about the right contour in the path integral 
for a pure cubic theory can be found in \cite{McKane1}) 
expanded in powers of $g$:

	\begin{equation}
            {\bf Z}_n(g)
	    \,=\,
	    \int\,{\cal D}\phi\,
	    e^{-\cal H}
	    \equiv
	    \sum_{K=0}^{\infty}\,Z_K(n)
	    \,g^{2K}
	    \label{expansion}
	\end{equation} 
We are interested in the asymptotic behaviour of $Z_K(n)$ as 
$K$ goes to infinity. We can extract $Z_K(n)$ from (\ref{expansion}) 
by the usual dispersion relation \cite{Kleinert,LGZJ}

	\begin{equation}
           Z_K(n)
	   \,=\,
 	   \int\,{\cal D}\phi\,
	   \frac{1}{2\pi i}
	   \,\oint\,dg^{2}\,\frac{1}{g^{2K}}
	   e^{- {\cal H}[\phi_i]}
	\end{equation} 
where the closed integral is performed in the cut complex plane $g^2$. 
In the large $K$ limit, the previous integral can be evaluated 
by the steepest descent 
method in the space $(\phi_i,g)$ and the saddle point equations are

	\begin{equation}
	   \nabla^2\phi_i
	   \,=\,
	   \frac{1}{2}\,
	   \epsilon_{ijk}\phi_j\phi_k
	   \label{first_equation}
	\end{equation}

	\begin{equation}
	   \frac{1}{3!}\,
	   \int\,d^dx\,
 	   \epsilon_{ijk}\phi_i\phi_j\phi_k
	   \,=\,
	   -\,
	   \frac{2K}{g}
	   \label{second_equation}
	\end{equation} 

Equations (\ref{first_equation}) and (\ref{second_equation}) 
can be decoupled easily by setting 

	\begin{eqnarray}
           \Phi_i=\frac{1}{2} \phi_i
	   &\,&
	   \Phi_i
	   = u_i \phi_c(x)
	   \label{saddle_point_equation}
	\end{eqnarray}

with the $u_i$'s satisfying 
	\begin{equation}
	   u_i
	   \,=\,
	   \epsilon_{ijk} u_j u_k.
	   \label{triangle}
	\end{equation} 
Equation (\ref{triangle}) has $n$ solutions $u_i^{(r)}$, with  
$r=1,2,...,n$, 

	\begin{equation} 
           u_i^{(r)}
	   \,=\, 
	   a^{(r)}\,\sum_{i=1}^r\,\epsilon_i^{\alpha}
	   \label{solutions}
	\end{equation} 
where 

	\begin{equation}
	   a^{(r)}
	   \,=\,
	   \frac{1}{(n+1)(n+1-2r)}
	\end{equation}

Working  in  $d=6$ dimension, the saddle point equations 
(\ref{first_equation},\ref{second_equation}) then read

	\begin{equation}
	   \nabla^2\phi_c(x)
	   \,=\,
	   \phi_c^2(x)
	   \label{critical_equation}
	\end{equation} 

	\begin{equation}
	   \frac{2}{3}\,\sum_i u_i u_i 
	   \,
	   \int\,d^6 x\,
	   \phi_c^3 (x)
	   \,=\,
	   -\,K g^2
	   \label{critical_equation_2}
	\end{equation}

whose general solution is 

	\begin{equation}
           \phi_c(x)
	   \,=\,
	   -\,
	   \frac{
		  24\lambda^2
	        }
		{ \left[ 
          		\,\lambda^2(x-x_0)^2\,+\,1
	          \right]^2
	        }
	   \label{solution}
	 \end{equation} 
where $\lambda$ and $x_0$ are constant parameters.
From equation (\ref{critical_equation_2}, \ref{solution}), we obtain  

	\begin{equation}
   	   g^2
	   \,=\,
  	   3\pi^3 \frac{2^8}{5K} 
 	\min_{\{r\}}\left(\sum_i u_i^{(r)} u_i^{(r)}\right) 
	\end{equation} 
and therefore  
	\begin{equation}
	   Z_K(n)
	   \,\sim\,
	   K!\,
	   \left(
                 \,\frac{5}{3\cdot 2^8 \pi^3 
		\min_{\{r\}}\left(\sum_i u_i^{(r)} u_i^{(r)}\right)}
	   \right)^K
	   \label{large_order}
	\end{equation} 
For the $(n+1)$ state Potts model ($n>1$ integer) the minimum  in 
(\ref{large_order}) 
is for $r=1$, giving,  as  expected 
for a pure cubic  theory,  a malign behaviour for the 
coefficients $Z_K$ which do not oscillate\cite{Kleinert,McKane1}

	\begin{equation}
	  Z_K(n)
	  \,\sim\,
	  K!\,
	  \left(
		\frac{5(n+1)^2 (n-1)^2}
		     {3\cdot 2^8 \pi^3\,n}
	 \right)^K
	 \label{high_order}
	\end{equation} 
  
For $n=0$ (corresponding to percolation problem) 
the saddle point cannot be obtained by a naive analytic 
continuation  of (\ref{large_order}) because evidently $r=1$ is not 
the dominating saddle point.

HRW\cite{HRW} found an alternative formulation for the percolation  
problem and they pointed out that in the $n\to 0$ limit 
the Feynman rules are the same as those of
the hamiltonian with two scalar fields

	\[
	   {\cal H}(\phi,\psi,g)
		\,=\,
	   \int\,d^6 x\,
		\left[
		       \frac{1}{2}(\nabla\phi)^2
			\,-\,
			\frac{1}{2}(\nabla\psi)^2
			\,+\,
			\frac{1}{2}(\phi^2-\psi^2)
			\,+\,
		\right.
	\]
	\begin{equation}
	   \hspace{1truecm}
	   +\,
	   \left.
 		 \frac{1}{3!}\,g\,
		 (\phi\,+\,\psi)^3
	   \right]
	   \label{Hamiltonian}
	\end{equation} 

provided that one keeps only $\phi-$connected diagrams as  those 
shown in figure 1. By means of path integrals techniques, they found 
the  coefficients  of the expansion 
of the two-point correlation function

	\begin{eqnarray}
             G(x,y)
		=
	     \sum_{K=0}^{\infty}
             \,G_K\,g^{2K}   &\,\,\,&
	   G_K \sim x K!\,
		\left( 
	              -\frac{15}{2^9 \pi^3}
		\right)^K
	\end{eqnarray}
which fortunately show an oscillating behaviour.

\section{HRW method for arbitrary $\lowercase{n}$}

As mentioned in the introduction, it would be useful to generalise the 
previous treatment to each  value of $n$.
Our first task is to write down the equivalent hamiltonian in analogy 
with  (\ref{Hamiltonian}).
In order to gain some insight,  let us compute the two-point 
correlation function up to fourth order 

	\begin{equation}
	  \langle
	        \phi_{\rm ext} 
		\rho_{ijk} \phi_i \phi_j \phi_k  
		\rho_{lmn} \phi_l \phi_m \phi_n 
		\rho_{abc} \phi_a \phi_b \phi_c 
		\rho_{def} \phi_d \phi_e \phi_f 
	        \phi_{\rm ext}
          \rangle
	\end{equation} 
by performing all the possible Wick contractions.  Because of the 
properties (\ref{epsilon_formulas}), we can get terms as

	\[
	   \sum_{\alpha,\beta,\gamma,\delta=1}^{n+1}
	   \,
	   \epsilon_i^{\alpha}
           \,
	   [(n+1)\delta^{\alpha\gamma}-1]
	   [(n+1)\delta^{\alpha\beta}-1]
           [(n+1)\delta^{\beta\gamma}-1]
	   \times\hfill
	\]
	\begin{equation}
	   \times\,
           [(n+1)\delta^{\beta\delta}-1]
	   [(n+1)\delta^{\gamma\delta}-1]
	   \,
	   \epsilon_f^{\delta}
	   \label{Wick_expansion2}
	\end{equation} 
Inspired by the previous discussion, we try to interpret 
$(n+1)\delta^{\alpha \beta}$ and $(-1)$ as propagators of two different  
(possibly multi-component) scalar fields, $\phi$ and $\psi$ respectively.
The several terms we obtain by expanding the products  in 
(\ref{Wick_expansion2}) correspond in the second theory to all 
the different diagrams with the same topology as the original ones. 
As in the $n \to 0$ case only diagrams with 
external $\phi$ legs connected by a $\phi$'s path do contribute.

\vspace{0.5cm}
\begin{center}
	{\large ** Here Figure 1 **}
\end{center}
\vspace{0.5cm}

In figure (1) we have drawn three of the possible graph built by 
$\phi$ and $\psi$ propagators. 
By requiring that the diagrams in the two theories have the same 
numerical coefficients, we obtain the Feynman rules.

Specifically, from the term involving only $\phi$ fields (figure (1a)) 
we get 
 
	\[
          \sum_{\alpha,\beta,\gamma,\delta=1}^{n+1}
	  \,
	  \epsilon_i^{\alpha}
	  \,
	  (n+1)^5
	  \delta^{\alpha\gamma}
          \delta^{\alpha\beta}
  	  \delta^{\beta\gamma}
          \delta^{\beta\delta}
  	  \delta^{\gamma\delta}
	  \,
  	  \epsilon_f^{\delta}=
	\]  
        \begin{equation}
	   =
	   (n+1)^5\,
	   \sum_{\alpha=1}^{n+1}\,
	   \epsilon_i^{\alpha}
	   \epsilon_j^{\alpha}
	 \end{equation} 
In these case, the presence of $\delta$'s eliminate the sum over the
four indices.  
If we associate a factor $(n+1)$ to each $\phi$ propagator, we obtain 
the same results without introducing any weight for the vertices. 

Let us now consider the second graph (see figure (1b)) 

	\[
	   \sum_{\alpha,\beta,\gamma,\delta}^{n+1}
	   \,
	   \epsilon_i^{\alpha}
	   \,(n+1)^4
	   \delta^{\alpha\beta}
	   \delta^{\beta\delta}
	   \delta^{\beta\gamma}
	   \delta^{\alpha\beta}
	   (-1)
           \,
	   \epsilon_f^{\delta}
	   \,=
	\]
	\begin{equation}
	   =\,
	   (-1)
	   (n+1)^4
	   \sum_{\alpha=1}^{n+1}\,
	        \epsilon_i^{\alpha}
	    	\epsilon_f^{\alpha}
	\end{equation} 
as we said, we associated a factor $(-1)$ for the $\psi$ propagator 
and we do not need any factor for the vertex $\phi^2 \psi$.

The situation is quite different when we deal with graphs where we 
have a $\psi^3$ vertex as in figure(1c) which derives from

	\[
	  (-1)
	  (-1)
	  (-1)
	  \sum_{\alpha,\beta,\gamma,\delta=1}^{n+1}
	  \,
	  \epsilon_i^{\alpha}
	  \,
	  (n+1)^2
	  \delta^{\alpha\beta}
	  \delta^{\beta\delta}
	  \,
	  \epsilon_f^{\delta}
	  \,=
	\]
	\begin{equation}
	  =\,
	  (-1)\,
	  (n+1)^2
	  \,
	  \sum_{\gamma=1}^{n+1}
	  \,
	  \left(
		\sum_{\alpha=1}^{n+1}
	        \,
		\epsilon_i^{\alpha}
		\epsilon_f^{\alpha}
	  \right).
	\end{equation} 

We note that we are left with a sum over $\gamma$ 
which gives a factor $(n+1)$. So a $\psi^3$ vertex requires 
a factor of $(n+1)$.  

At higher order we also have non-zero diagrams with also
$\phi$ connected sub-graphs 
as shown in figure (2)

\vspace{0.5cm}
\begin{center}
	{\large ** Here Figure 2 **}
\end{center}
\vspace{0.5cm}

It can be rather simply proven that each $\phi-$connected subgraph
gives a factor of $(n+1)$.
Summarising the Feynman rules for $(n+1)$ component Potts
model are 

\begin{itemize}
  \item[1)]{Only $\phi-$connected graphs give a non zero contribution.}
  \item[2)]{Each $\phi$ propagator gives a factor of $(n+1)$.}
  \item[3)]{Each $\psi$ propagator  gives a factor of $(-1)$.}
  \item[4)]{Each $\psi^3 $vertex gives a factor of $(n+1)$.}
  \item[5)]{Each $\phi-$connected subgraph  gives a factor of $(n+1)$.}
\end{itemize}

These Feynman rules correspond to the following hamiltonian 

	\[
	   {\cal H}\,=\,
	   \int\,d^6 x\,
		\left\{
			\frac{1}{2}
			\sum_{\alpha=1}^{n+1}
			\left(
			      \frac{1}{n+1}
			      (\nabla \phi_{\alpha})^2
			      \,+\,
			      \frac{1}{n+1}
			      \phi_{\alpha}^2
			\right)\,+
		\right.\hfill
	\]
	\begin{equation}
		\left.
		-\,
		      \left(
			    \frac{1}{2}(\nabla \psi)^2
			    \,+\,
			    \frac{1}{2}\psi^2
		       \right)
			\,+\,
			\frac{g}{3!}\,
			\sum_{\alpha=1}^{n+1}\,
			(\phi_{\alpha}\,+\,\psi)^3
		\right\}
		\label{alternative_Hamiltonian} 
	\end{equation} 

By following exactly the same procedure as HRW's in order to extract 
the $\phi-$connected diagrams, we get for the two-point correlation 
function

	\[
	   G_{\phi-c}^{(2)}(x,y)
	   \,=\,
	   \frac{
		  \int\,{\cal D}\phi\,\phi(x)\phi(y)\,\exp[-{\cal H}^*(\phi)] 
                  \,e^{n{\cal W}^*}
		}
	        {
		  \int\,{\cal D}\psi \left[e^{{\cal W}^*}\right]^{n+1}
		}
		\,+\,
	\]
	\[
	   -\,
	   \int\,{\cal D}\psi{\cal D}\phi^{(1)}{\cal D}\phi^{(2)}
	   \,\phi^{(1)}(x)\phi^{(2)}(y)\,
	   \times
	   \hfill
	\]
	\begin{equation}
	   \times
	    \frac{
		   \exp\left[ -{\cal H}^*(\phi^{(1)})-{\cal H}^*(\phi^{(2)})
                       \right]
		 }
		 {
			  e^{(1-n){\cal W}^*}\,
			  \,
			   \left[ \int\,{\cal D}\psi
				  \,e^{(n+1){\cal W}*}
			   \right]
		 }
	    \label{final_twopoint_Green}
	\end{equation} 

where ${\cal H}^*(\phi)$ is the reduced hamiltonian 

	\begin{eqnarray}
	  {\cal H}^*(\phi,\psi) 
	  =
	  \int dx^6
	  \left\{
		 \frac{1}{2}\frac{1}{n+1} 
	         \left( (\nabla\phi)^2 + \phi^2 \right)
	  \right. 
		 &-&
		 \frac{1}{2}\frac{1}{n+1} 
	         \left( (\nabla\psi)^2 + \psi^2 \right) + 
	      \nonumber \\[0.5cm] 
	&+&
	\left.
	     \frac{g}{3!}\left(\phi+\psi\right)^3
	\right\}
	\end{eqnarray}
and ${\cal W}^*(\psi)$ is defined by 

	\begin{equation}
	  \exp\left[{\cal W}^*(\psi)\right]
	  =
	  \int {\cal D} \phi \exp-{\cal H}^*(\phi,\psi)
	\end{equation}

\section{ High order behaviour \label{Asympt}}

As in the $n\to 0$  case, it can be shown that only the second term 
on the left hand side of 
(\ref{final_twopoint_Green}) contributes to the high order behaviour of the 
perturbative expansion because of the particular form of the cubic 
interaction and we have to evaluate 

          \[
	   -\,
	   \int\,{\cal D}\psi{\cal D}\phi^{(1)}{\cal D}\phi^{(2)}
	   \,\phi^{(1)}(x)\phi^{(2)}(y)\,
	   \frac{1}{2\pi i}\,\oint dg^2\, \frac{1}{g^2K}\,
	   \times
	   \hfill
	\]
	\begin{equation}
	   \times
	    \frac{
		   \exp\left[ -{\cal H}^*(\phi^{(1)})-{\cal H}^*(\phi^{(2)})
                       \,+\,2K\log g \right]
		 }
		 {
			  e^{(1-n){\cal W}^*}\,
			  \,
			   \left[ \int\,{\cal D}\psi
				  \,e^{(n+1){\cal W}*}
			   \right]
		 }
	\end{equation} 

for large $K$. Our goal can be achieved as usual  
by the steepest descent method.
The equations for the saddle points are

\begin{eqnarray}
       \frac{1}{n+1} \nabla ^2 \phi_i  &=& 
                         -\frac{1}{2} g\,(\psi+\phi_i)^2   
          \hspace{0.5cm} i=1,2\\[0.3cm]     
          \frac{2}{n+1}\nabla^2 \psi &+&
                 \frac{1}{2}\,g\,(\psi+\phi_1)^2
			    \,+\frac{1}{2}\,g\,(\psi+\phi_2)^2+
			    (1-n)
	\frac{\partial {\cal W}^*}{\partial\psi}\,=\,0  
\end{eqnarray}
\begin{equation}
  \frac{1}{3!}\int d^6 x 
	      \left\{
                     (\psi+\phi_1)^3\,+\,(\psi+\phi_2)^3
	      \right\}   
		\,+\,(1-n)
		\frac{\partial {\cal W}^*}{\partial g}\,+\,
		\frac{2K}{g} = 0
\end{equation}
It is simple to evaluate the partial derivative of ${\cal W}^*$ 
with respect to $g$ 
and $\psi$ by steepest descent method too and we obtain 

	  \[
	    \frac{\partial {\cal W}^*}{\partial \psi}
	    \,=\,
	    \left(
                   \,-\frac{1}{n+1} \nabla^2 \psi
                   \,-\,
                   \frac{1}{2} g\,(\phi_0\,+\,\psi)^2\,
            \right)
	    \left[1\,+\,O(g^2)\right]
	  \]\par\noindent
and
	  \[
	    \frac{\partial {\cal W}^*}{\partial g}
	    \,=\,
	    -\frac{1}{3!}\,
	    \left(
                  \int d^6 x (\psi\,+\,\phi_0)^3
             \right)
	    \left[1\,+\,O(g^2)\right]
	  \]\par\noindent
provided that $\phi_0$ obeys to the following equation 

	\[
	  \frac{1}{n+1} \nabla^2 \phi_0\,
	   \,=\, 
	   \frac{1}{2} g\,(\phi_0\,+\,\psi)^2
	\]\par\noindent
After the  re-scaling   

	\[
	  \phi_i\longrightarrow\Phi_i\,=\,\frac{1}{g}\,\phi_i
	  \,\,\, i=0,1,2
	\]\par\noindent
	\[
	  \psi\longrightarrow\Psi\,=\,\frac{1}{g}\,\psi
	\]\par\noindent

the saddle point equation are finally  

	\begin{equation}
	   \frac{1}{n+1} \Phi_i
             \,=\,
            \frac{1}{2}
            \left(\Phi_i\,+\,\Psi\right)
	   \,\,\, i=0,1,2
	   \label{equation_1_2_3}
	\end{equation} \par\noindent
	\begin{equation}
	   \nabla^2 \Psi
            \,+\,
            \frac{1}{2} \left(\Phi_1\,+\,\Psi\right)^2
             \,+\, 
	    \frac{1}{2} \left(\Phi_2\,+\,\Psi\right)^2
            \,-\, 
            \frac{(1-n)}{2}\left(\Phi_0\,+\,\Psi\right)^2 
	   \label{equation_4}         
	\end{equation} \par\noindent   
	\begin{equation}
	  -2\,K\,g^2\,=\,\frac{1}{3!}
                      \int  d^6 x \,
                      \left\{ 
                         \left(\Phi_1\,+\,\Psi\right)^3\,+\,
			 \left(\Phi_2\,+\,\Psi\right)^3\,-\,
			 (1-n)\left(\Phi_0\,+\,\Psi\right)^3
		      \right\}
	   \label{equation_5}
	\end{equation} \par\noindent
In order to solve the equations (\ref{equation_1_2_3},\ref{equation_4},
\ref{equation_5}),  let us make the ansatz  

	\begin{equation}
           \Phi_i\,=\,u_i\,\phi_c(x)
	   \hspace{0.5truecm} 
	   \Psi\,=\,\nu \, \phi_c(x)
	\end{equation} 
where $\phi_c(x)$ is the solution of (\ref{critical_equation}).
Therefore, we have reduced our differential equations to algebraic ones.

\begin{eqnarray*}
     && \frac{1}{(n+1)}\,u_i = \frac{1}{2} (\nu\,+\,u_i)^2  \\[0.5cm]
     && \nu = -\frac{1}{(n+1)}\,u_1
               -\frac{1}{(n+1)}\,u_2
	       +\frac{(1-n)}{(n+1)}\,u_0 
	     \\[0.5cm]
     && K g^2 =
		 \frac{3\cdot2^6\pi^3}{5}
                 \left[ 
		       (\nu+u_1)^3+(\nu+u_2)^3+(n-1)(\nu+u_0)^3
		 \right] 
\end{eqnarray*}	 
Solving the first equation gives two solution 

	\begin{equation}
            u^{(\pm)}
	      \,=\,
	    -\left(
                   \nu-\frac{1}{(n+1)}
	     \right)
	     \pm
	     \frac{1}{(n+1)}
	 	\left[
		      1-2\nu(n+1)
		\right]^{\frac{1}{2}}
	\end{equation} 
Only four choices are possible 

\begin{itemize}
\item[1)]{$u_1=u^{(+)} \,\,\, u_2=u_0=u^{(-)}$} 
\item[2)]{$u_1=u^{(-)} \,\,\, u_2=u_0=u^{(+)}$}
\item[3)]{$u_0=u^{(+)} \,\,\, u_1=u_2=u^{(-)}$}
\item[4)]{$u_0=u^{(-)} \,\,\, u_1=u_2=u^{(+)}$}
\end{itemize}

From the $1)$ and $2)$ we obtain two equivalent solutions provided 
that $n\neq 1$ 

	\begin{equation}
	   \nu
            \,=\,
	   -\frac{2n}{(n+1)\,(n-1)^2}
	    \hspace{1.5truecm}n \neq 1
	\end{equation} 
The choices $3)$ and $4)$ gives other two equivalent solution for $\nu$ 
when $n\neq 3$
	
	\begin{equation}
	  \nu
	  \,=\,
	  \frac{4 (1-n)}{(n+1)\,(n-3)^2}
	   \hspace{1.5truecm}n \neq 3                   
	\end{equation} 
The previous two value  for $\nu$ give two saddle points for $g^2$

	\begin{equation}
	   g_{(1)}^2
	   \,=\,
	   \,\frac{1}{K}\,
	   \frac{3\cdot 2^8 \pi^3}{5}\,
	   \frac{n}{(n+1)^2\,(n-1)^2}
	   \hspace{1.0truecm}n\neq 1
	\end{equation} 

and

	\begin{equation}
	   g_{(2)}^2
	   \,=\,
	   \,\frac{1}{K}\,
	   \frac{3\cdot 2^9 \pi^3}{5}\,
	   \frac{(n-1)}{(n+1)\,(n-3)^2}
	   \hspace{1.0truecm}n\neq 3
	\end{equation} 

A this point, it is interesting to note that the saddle points 
$g^2_{(1)}$ and $g^2_{(2)}$ are respectively the $r=1$ and the $r=2$ 
points of (\ref{solution}). Anyway, 
Now, our final aim  is to evaluate the large order behaviour of the 
perturbative expansion, that is the coefficients $Z_K$ in the expression 
(\ref{expansion}). So a little care needs to understand which saddle 
point dominates at $n$ fixed. 
We can obtain the two following behaviours

	\begin{equation}
	   Z^{(1)}_K
           \,\sim\,  	
	   \left(
                 \,\frac{5}{3\cdot2^8 \pi^3}\,
	         \frac{(n+1)^2\,(n-1)^2}{n}\,
	   \right)^K
	   \hspace{1truecm}n \neq 1
	   \label{first_behaviour}
	\end{equation} 
and

	\begin{equation}
	  Z^{(2)}_K
	  \,\sim\,
	    \left(
	          \,\frac{5}{3\cdot 2^9 \pi^3}\,
		  \frac{(n+1)^2\,(n-3)^2}{(n-1)}
	    \right)^K
	    \hspace{1truecm} n \neq 3	
	    \label{second_behaviour}
	\end{equation} 
For $n=0$ (percolation) the second saddle point dominate over 
the first one and we recover the HRW results. Actually there exist 
other formulations for the percolation problem disagreeing with HRW 
\cite{McKane2}, but it can be shown\cite{Caracciolo} that the different 
result was due to an unappropriate choice of an integration contour. 
Nevertheless, the nature of the percolation behaviour  still remains  
controversial because there is another 
result\cite{McKane3} where the $r=1$ 
saddle point seems to dominate.    
For $n=1$ (Ising model) we cannot choose the coefficient $Z^{(1)}_K$
and $Z^{(2)}_K$  is identically zero according to the fact that 
the trilinear interaction vanishes as $n=1$. 
For $n$ integer we have to take into account the (\ref{first_behaviour})
and this is in agreement with the calculations  performed in section 
(\ref{Potts_model})

\section*{Conclusions}

We performed the computation of the large order behaviour of the 
perturbative expansion of the $(n+1)$ state Potts model with a cubic 
interaction for a generic value of $n$, by mapping the original problem 
in a simpler one. We showed that our equivalent field theory 
has two saddle points depending on $n$ 
and one has to choose the smallest one in order 
to extract the dominant term at large order. 
A natural question can arise 
looking at the different number of the saddle points in both the 
formulations. Indeed in Sec.I it has been shown that the Potts 
model, when studied by the usual techniques, has $n$ saddle points 
characterised by an integer $r$, whereas our alternative 
formulation gives only two inequivalent saddle points. 
Indeed, the large order behaviour is related to the cut 
discontinuity of the correlation function for small value 
of the coupling constant, that is the imaginary part of the 
Green functions generated by the Taylor expansions around the 
saddle point. 
For $n=2,3,...$ the imaginary part of the correlation functions is 
dominated by the $r=1$ saddle point. When an analytic continuation 
in $n$ is performed and some particular limit are taken , 
the $r=2$ saddle point becomes the leading term.
All the other saddle points are sub-leading and they do not affect the 
large order behaviour. This generalisation can help to compute 
critical exponents for theory where  $n$ is not integer as those 
mentioned in the introduction eventually by adding higher order 
interactions.

\section*{Acknowledgements}

It is a great pleasure to thank Giorgio Parisi for having suggested 
and supervised this work and G.F.Bonini for scientific discussions.

\clearpage

\thispagestyle{empty}

\section*{List of captions}

\begin{itemize}

\item[Figure 1: ]{Some diagrams built by means of $\psi$ (solid line) 
		  and $\psi$ (dotted line) propagators.
		  Only the diagrams $(a)$ and $(b)$ are suitable because 
	 	     $\phi$ connected.}

\item[Figure 2: ]{High order diagrams with a $\phi$ connected 
                  subgraph. The filled region means any non-zero diagram.}
\end{itemize}

\clearpage

\thispagestyle{empty}

	\begin{figure}
            \centertexdraw{
                   \drawdim cm  \linewd 0.05
                    \move(-6 0)
                    \lvec (-5 0)
                    \move (-4 0)
                    \lcir r:1
                    \move(-3 0)
                    \lvec (-2 0)
                     \move (-4 -1)
                    \lvec (-4 1)
                    \htext(-4.5 -2){\Large {\bf (a)}}
                    \move (2 0)
                    \lvec (3 0)
                    \move (4 0)
                    \larc r:1 sd:180 ed:360
                    \lpatt (0.1 0.1) 
                    \larc r:1 sd:0 ed:180
                    \move (5 0)
                   \lpatt() \lvec(6 0)
                   \move (4 -1)
                   \lvec (4 1)
                   \htext(3.5 -2){\Large {\bf (b)}}
                   \move(0 -4)
                    \lpatt(0.1 0.1) 
                    \larc r:1 sd:0 ed:270 
                   \lpatt() \larc r:1 sd:270 ed:360
                   \move(0 -5)
                   \lvec(0 -3)
                   \move (-2 -4)
                   \lvec (-1 -4)
                   \move (1 -4)
                   \lvec (2 -4)
                   \htext(-0.5 -6){\Large {\bf (c)}}
                 }
	    \caption{}
	    \label{diagrams}
	\end{figure}

\clearpage 
\thispagestyle{empty}

	\begin{figure}
             \centertexdraw{
                   \drawdim cm  \linewd 0.05
                    \move (-4 -6)
                    \lvec (-3 -6)
                    \move (-3 -6.5)
	            \lvec (-3 -5.5)
		    \lvec (4 -5.5)
	            \lvec (4 -6.5)
		    \lvec (-3 -6.5)
	            \lfill f:0.3
		    \move(4 -6)
		    \lvec(5 -6)
		    \lpatt(0.1 0.1)
		    \move(-1 -5.5)
		    \larc r:1 sd:0  ed:180
		    \move(2 -5.5)
		    \larc r:1 sd:0 ed:180
		    \move(0.5 -5.5)
		    \larc r:1 sd:45 ed:135
		    \lpatt( )
		    \move(0 -4.5)
		    \larc r:0.5 sd:0 ed:185
		    \move(2 -4.5)
		    \larc r:0.5 sd:-10 ed:195
		    \move(1.5 -4.5)
		    \larc r:0.5 sd:60 ed:200
		    \move(0.6 -4.5)
		    \larc r:0.5 sd:30 ed:130
                        }		    
	   \caption{}
	    \label{f4}
	\end{figure}


\begin{thebibliography}{99}

\bibitem{Potts} 
R.~B.~Potts, Proc. Camb. Phil. Soc. {\bf 48}, 106  (1952). 

\bibitem{Ising} 
E.~Ising,  Z. Phys.  {\bf 21},  613 (1925).       

\bibitem{Domb}  
C.~Domb, J. Phys A  {\bf 7},  1335 (1974).       

\bibitem{Kasteleyn} 
P.~W.~Kasteleyn and C~.M.~Fortuin, 
 J. Phys. Soc. Jap. (Suppl.) {\bf 26}, 11  (1969).       

\bibitem{Giri} 
M.~R.~Giri, M.~J.~Stephen and  G.~S.~Grest, 
Phys. Rev. B {\bf 16},  4971 (1977).       

\bibitem{Kunz} 
H.~Kunz and F.~Y.~Wu, J. Phys. C  {\bf 11},  L1 (1978).      

\bibitem{Fortuin} 
C.~M.~Fortuin and P.~W.~Kasteleyn, Physica {\bf 57}, 536 (1972).

\bibitem{Kirchhoff} 
G.~Kirchhoff, Ann. Phys. Chem {\bf 72}, 497 (1847) 

\bibitem{Aharony1} 
A.~Aharony, J. Phys. C  {\bf 11},  L457 (1978).        

\bibitem{Aharony2} 
A.~Aharony and P.~Pfeuty, J. Phys. C  {\bf 12}, L125 (1979).       

\bibitem{Lubensky} 
T.~C.~Lubensky and J.~Isaacon, Phys. Rev. Lett. {\bf 41}, 829 (1978). 
      
\bibitem{Wu}
F.~Y.~Wu, Rev. Mod. Phys. {\bf 54}, 235 (1982).

\bibitem{Zia}
R.~K.~Zia and D.~Wallace, J. Phys. A {\bf 8} 1495 (1975).


\bibitem{HRW}
A.~Houghton, J.~S.~Reeve and D.~J.~Wallace, Phys.Rev. B {\bf 17}, 2956 (1978). 


\bibitem{McKane1}
A.~J.~McKane, Nucl. Phys. B {\bf 152} 166 (1979). 


\bibitem{Kleinert}
H.~Kleinert, {\sl Path Integrals in Quantum Mechanics, Statistics and 
0Polymer Physics}, (World Scientific 1990).


\bibitem{LGZJ}
J.~C.~Le Guillou and J.~Zinn Justin eds., {\sl Large Order Behaviour of 
Perturbation
Theory}, (North-Holland, 1990).


\bibitem{McKane2}
A.J.McKane, {\sl J. Phys. A} {\bf 19} 453 (1986)


\bibitem{Caracciolo} 
S.Caracciolo and D.M.Carlucci, {\sl to be published}


\bibitem{McKane3}
E.Br\'ezin and A.J.McKane, {\sl J. Phys. A} {\bf 20} 2235 (1987) 


\end{thebibliography}
\end{document}